\pdfoutput=1
\documentclass{vldb}

\usepackage{graphicx}
\usepackage{balance}
\usepackage{times}

\newcommand{\subparagraph}{\paragraph}
\usepackage{titlesec}
\titleformat{\paragraph}[runin]
    {\normalfont\normalsize\bfseries}{\theparagraph}{1em}{}

\titleformat{\subsubsection}[runin]
        {\normalfont\bfseries}
        {\thesubsubsection}
        {0.5em}
        {}
        [. ]

\usepackage{booktabs} 

\usepackage{xcolor}
\usepackage[utf8]{inputenc}
\usepackage{xspace}
\usepackage{subcaption}
\usepackage{url}
\usepackage{breakurl}

\usepackage{enumitem}
\usepackage[abbreviations,binary-units]{siunitx}
\usepackage{amsthm}
\usepackage{glossaries}
\glsdisablehyper
\setacronymstyle{long-short}

\newacronym{ccdf}{CCDF}{complementary cumulative distribution function}
\newacronym{rss}{RSS}{resident set size}

\usepackage{tikz}
\usetikzlibrary{backgrounds,fit,positioning}
\usepackage{tkz-graph}

\usepackage[final]{changes}

\usepackage{listings}
\theoremstyle{definition}
\newtheorem{definition}{Definition}
\newtheorem{property}{Property}

\usepackage[activate={true,nocompatibility},final,tracking=true,kerning=true,spacing=true,factor=1100,stretch=10,shrink=10]{microtype}
\usepackage[capitalize,noabbrev]{cleveref}

\newcommand{\pname}{Megaphone\xspace}
\newcommand{\sys}{\pname}

\newcommand{\fixerop}{\ensuremath{F}\xspace}
\newcommand{\stuffop}{\ensuremath{S}\xspace}
\newcommand{\logicop}{\ensuremath{L}\xspace}

\usepackage{bold-extra}

\newcommand{\stitle}[1]{\paragraph{#1}}

\newcounter{VasiaNOC}
\stepcounter{VasiaNOC}

\newcounter{JohnNOC}
\stepcounter{JohnNOC}

\newcounter{MoritzNOC}
\stepcounter{MoritzNOC}

\vldbTitle{\pname: Live state migration for distributed streaming dataflows}
\vldbAuthors{Moritz Hoffmann, Andrea Lattuada, Frank McSherry, Vasiliki
Kalavri, John Liagouris, and Timothy Roscoe}
\vldbDOI{https://doi.org/10.14778/xxxxxxx.xxxxxxx}
\vldbVolume{12}
\vldbNumber{xxx}
\vldbYear{2019}

\begin{document}

\title{\pname: Latency-conscious state migration for distributed streaming dataflows}
\numberofauthors{3}
\author{
\alignauthor
Moritz Hoffmann
\alignauthor
Andrea Lattuada
\alignauthor
Frank McSherry
\and
\alignauthor
Vasiliki Kalavri
\alignauthor
John Liagouris
\alignauthor
Timothy Roscoe
\and \alignauthor
\affaddr{Systems Group, ETH Zurich}
\email{first.last@inf.ethz.ch}
}
\additionalauthors{}
\date{}
\maketitle

\begin{abstract}

We design and implement \pname, a data migration mechanism for stateful distributed dataflow engines with latency objectives.
When compared to existing migration mechanisms, \pname has the following differentiating characteristics: (i) migrations can be subdivided to a configurable granularity to avoid latency spikes, and (ii) migrations can be prepared ahead of time to avoid runtime coordination.
\pname is implemented as a library on an unmodified timely dataflow implementation, and provides an operator interface compatible with its existing APIs.
We evaluate \pname on established benchmarks with varying amounts of state and observe that compared to na\"ive approaches \pname reduces service latencies during reconfiguration by orders of magnitude without significantly increasing steady-state overhead.

\end{abstract}





\section{Introduction}\label{sec:introduction}
Distributed stream processing jobs are long-running dataflows that continually ingest data from sources with dynamic rates and must produce timely results under variable workload conditions~\cite{pokemon-case,twitter-case}.

To satisfy latency and availability requirements, modern stream processors support \emph{consistent online reconfiguration}, in which they update parts of a dataflow computation without disrupting its execution or affecting its correctness.
Such reconfiguration is required during \textbf{rescaling} to handle increased input rates or reduce operational costs~\cite{Fernandez,Floratou2017}, to provide \textbf{performance isolation} across different dataflows by dynamically scheduling queries to available workers, to allow \textbf{code updates} to fix bugs or improve business logic~\cite{Armbrust18,Carbone17}, and to enable \textbf{runtime optimizations} like execution plan switching~\cite{Zhu04} and straggler and skew mitigation~\cite{Fang17}.

\begin{figure}[t]
    \centering
    {\includegraphics[width=.99\linewidth,page=3,trim={0 6pt 0 12pt}]{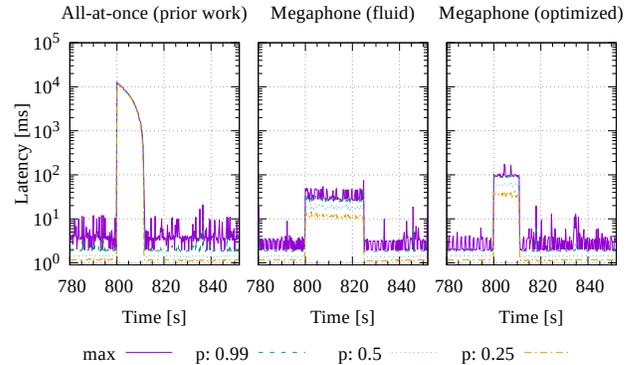}}

    \caption{A comparison of service latencies in prior coarse-grained migration strategies (all-at-once) with two of \pname's fine-grained migration strategies (fluid and optimized), for a workload that migrates one billion keys consisting of 8GB of data.}


    \label{fig:wc_migration_timeline}
\end{figure}

Streaming dataflow operators are often stateful, partitioned across workers by key, and their reconfiguration requires \emph{state migration}: intermediate results and data structures must be moved from one set of workers to another, often across a network.
Existing state migration mechanisms for stream processors either pause and resume parts of the dataflow (as in Flink~\cite{flink}, Dhalion~\cite{Floratou2017}, and SEEP~\cite{Fernandez}) or launch new dataflows alongside the old configuration (as for example in ChronoStream~\cite{Wu2015} and Gloss~\cite{Rajadurai18}).  In both cases state moves ``all-at-once,'' with either high latency or resource usage during the migration.


State migration has been extensively studied for distributed databases~\cite{Barker2012,Das2011,Elmore2011,Elmore2015}.
Notably, Squall~\cite{Elmore2015} uses transactions to multiplex fine-grained state migration with data processing.
These techniques are appealing in spirit, but use mechanisms (transactions, locking) not available in high-throughput stream processors and are not directly applicable without significant performance degradation.

In this paper we present \pname, a technique for fine-grained migration in a stream processor which delivers maximum latencies orders of magnitude lower than existing techniques, based on the observation that a stream processor's structured computation and logical timestamps allow the system to \emph{plan} fine-grained migrations.
\pname can specify migrations on a key-by-key basis, and then optimizes this by batching at varying granularities; as Figure~\ref{fig:wc_migration_timeline} shows, the improvement over all-at-once migration can be dramatic.
\added{
This paper is an extended version of a preliminary workshop publication}~\cite{hoffmann2018latency}\added{.
In this paper, we describe a more general mechanism, further detail its implementation, and evaluate it more thoroughly on realistic workloads.
}


Our main contribution is \emph{fluid migration} for stateful streaming
dataflows: a state migration technique that enables consistent online
reconfiguration of streaming dataflows and smoothens latency spikes
without using additional resources (\cref{sec:mechanism}) by employing
fine-grained planning and coordination through logical timestamps.
Additionally, we design and implement an API for reconfigurable stateful
timely dataflow operators that enables fluid migration to be
controlled simply through additional dataflow streams rather than through changes to the
dataflow runtime itself (\cref{sec:implementation}).
Finally, we show that \pname has negligible steady-state overhead and
enables fast direct state movement using the NEXMark benchmarks suite
and microbenchmarks (\cref{sec:evaluation}).

\pname is built on timely dataflow,\footnote{\scriptsize
  \url{https://github.com/frankmcsherry/timely-dataflow}} and is
implemented purely in library code, requiring no modifications to the
underlying system.  We first review existing state migration
techniques in streaming systems, which either cause performance degradation
or require resource overprovisioning. We also review live migration in
DBMSs and identify the technical challenges to implement similar
methods in distributed stream processors (\cref{sec:background}).

\section{Background and Motivation}\label{sec:background}


A distributed dataflow computation runs as a physical execution plan which maps operators to provisioned compute resources (or workers). The execution plan is a directed graph whose vertices are operator instances (each on a specific worker) and edges are data channels (within and across workers). Operators can be stateless (e.g., filter, map) or stateful (e.g., windows, rolling aggregates). State is commonly partitioned by key across operator instances so that computations can be executed in a data-parallel manner. At each point in time of a computation, each worker (with its associated operator instances) is responsible for a set of keys and their associated state.

\emph{State migration} is the process of changing the assignment of keys to workers and redistributing respective state accordingly.
A good state migration technique should be \emph{non-disruptive} (minimal increase in response latency during migration), \emph{short-lived} (migration completes within a short period of time), and \emph{resource-efficient} (minimal additional resources required during the migration).

We present an overview of existing state migration strategies in distributed streaming systems and identify their limitations. We then review live state migration methods adopted by database systems and provide an intuition into \pname's approach to bring such migration techniques to streaming dataflows.

\subsection{State migration in streaming systems}\label{sec:migration-strategies}
Distributed stream processors, including research prototypes and production-ready systems, use one of the following three state migration strategies.

\stitle{Stop-and-restart}
A straight-forward way to realize state migration is to temporarily stop program execution, safely transfer state when no computation is being performed, and restart the job once state redistribution is complete. This approach is most commonly enabled by leveraging existing fault-tolerance mechanisms in the system, such as global state checkpoints. It is adopted by Spark Streaming~\cite{zaharia2013discretized}, Structured Streaming~\cite{Armbrust18}, and Apache Flink~\cite{Carbone17}.

\stitle{Partial pause-and-resume}
In many reconfiguration scenarios only one or a small number of operators need
to migrate state, and halting the entire dataflow is usually
unnecessary. An optimization first introduced in Flux~\cite{Shah02} and later
adopted in variations by Seep~\cite{Fernandez}, IBM Streams~\cite{ibm_streams},
StreamCloud~\cite{Gulisano2012}, Chi~\cite{Mai18}, and FUGU~\cite{Heinze:2014:LES:2611286.2611294}, pauses the computation only for the affected dataflow subgraph. Operators not participating in the migration continue without interruption. This approach can use fault-tolerance checkpoints for state migration as in~\cite{Fernandez,Mai18} or state can be directly migrated between operators as in~\cite{Gulisano2012,Heinze:2014:LES:2611286.2611294}.

\stitle{Dataflow Replication}
To minimize performance penalties, some systems replicate the whole dataflow or subgraphs of it and execute the old and new configurations in parallel until migration is complete. ChronoStream~\cite{Wu2015} concurrently executes two or more computation slices and can migrate an arbitrary set of keys between instances of a single dataflow operator.
Gloss~\cite{Rajadurai18} follows a similar approach and gathers operator state during a migration in a centralized controller using an asynchronous protocol.


\medskip
Current systems fall short of implementing state migration in a non-disruptive and cost-efficient manner.
Existing stream processors migrate state \emph{all-at-once}, but differ in whether they pause the existing computation or start a concurrent computation.
As Figure~\ref{fig:wc_migration_timeline} shows, strategies that pause the computation can cause high latency spikes, especially when the state to be moved is large.
On the other hand, dataflow replication techniques reduce the interruption, but at the cost of high resource requirements and required support for input duplication and output de-duplication. For example, for ChronoStream to move from a configuration with $x$ instances to a new one with $y$ instances, $x+y$ instances are required during the migration.

\subsection{Live migration in database systems}\label{sec:migration_db}
Database systems have implemented optimizations that explicitly target limitations we have identified in the previous section, namely unavailability and resource requirements.
Even though streaming dataflow systems differ significantly from databases in terms of data organization, workload characteristics, latency requirements, and runtime execution, the fundamental challenges of state migration are common in both setups.

Albatross~\cite{Das2011} adopts VM live migration techniques and is further optimized in~\cite{Barker2012} with a dynamic throttling mechanism, which adapts the data transfer rate during migration so that tenants in the source node can always meet their SLOs. 
Prorea~\cite{Schiller2013} combines push-based migration of hot pages with pull-based migration of cold pages techniques.
Zephyr~\cite{Elmore2011} proposes a technique for live migration in shared-nothing transactional databases which introduces no system downtime and does not disrupt service for non-migrating tenants.

The most sophisticated approach is Squall~\cite{Elmore2015}, which interleaves state migration with transaction processing by (in part) using transaction mechanisms to effect the migration.
Squall introduces a number of interesting optimizations, such as pre-fetching and splitting reconfigurations to avoid contention on a single partition.
In the course of a migration, if migrating records are needed for processing but not yet available, Squall introduces a transaction to acquire the records (completing their migration).
This introduces latency along the critical path, and the transaction locking mechanisms can impede throughput, but the system is neither paused nor replicated. To the best of our knowledge, no stream processor implements such a fine-grained migration technique.

\subsection{Live migration for streaming dataflows}



Applying existing fine-grained live migration techniques to a streaming engine is non-trivial.
While systems like Squall target OLTP workloads with short-lived transactions, streaming jobs are long-running. In such a setting, Squall's approach to acquire a global lock during initialization is not a viable solution. Further, many of Squall's remedies are reactive rather than proactive (because it must support general transactions whose data needs are hard to anticipate), which can introduce significant latency on the critical path.

The core idea behind \pname's migration mechanism is to multiplex \emph{fine-grained} state migration with actual data processing, coordinated using logical timestamps common in stream processors. This is a proactive approach to migration that relies on the prescribed structure of streaming computations, and the ability of stream processors to coordinate with high frequency using logical timestamps. Such systems, including \pname, avoid the need for system-wide locks by pre-planning the rendezvous of data at specific workers.





\section{State migration design}\label{sec:mechanism}

\pname's features rely on \added{core streaming dataflow concepts such as logical time, progress tracking, data-parallel operators, and state management. Basic implementations of these concepts are present in all modern stream processors, such as Apache Flink}~\cite{flink}, \added{Millwheel}~\cite{Akidau}, \added{and Google Dataflow}~\cite{Akidau:2015:DMP:2824032.2824076}.
\added{In the following, we rely on the Naiad}~\cite{murray:naiad} \added{timely dataflow model as the basis to describe the \pname migration mechanism. Timely dataflow natively supports a superset of dataflow features found in other systems in their most general form.}


\subsection{Timely dataflow concepts}\label{sec:dataflow_concepts}

A streaming computation in Naiad is expressed as a \emph{timely dataflow}:
a directed (possibly cyclic) graph where nodes represent stateful operators and edges represent data streams between operators.
Each data record in timely dataflow bears a \emph{logical timestamp}, and operators maintain or possibly advance the timestamps of each record.
Example timestamps include integers representing milliseconds or transaction identifiers, but in general can be any set of opaque values for which a partial order is defined.
The timely dataflow system tracks the existence of timestamps, and reports as processed timestamps no longer exist in the dataflow, which indicates the forward progress of a streaming computation.

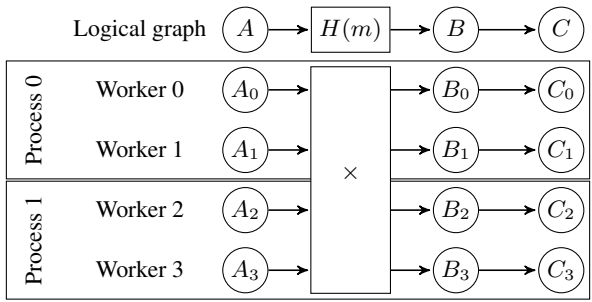
\begin{figure}[tbp]
    \centering
    \begin{tikzpicture}[
        node distance=.8cm and 1.4cm,on grid,auto,
        text height=1.5ex,text depth=.25ex,
        minimum height=6mm,
        op/.style={circle,draw,inner sep=1pt},
        hash/.style={rectangle,draw},
        phantom/.style={rectangle},
        exchange/.style={rectangle,draw,inner sep=0pt,fill=white},
        label/.style={},
        process/.style={rotate=90},
        ]
        \node[label] (LG) {Logical graph};
        \node[op] (A) [right=of LG] {$A$};
        \node[hash] (H) [right=of A] {$H(m)$}
            edge [pre] (A);
        \node[op] (B) [right=of H] {$B$}
            edge [pre] (H);
        \node[op] (C) [right=of B] {$C$}
            edge [pre] (B);

        \foreach \x/\b in {0/LG, 1/W 0, 2/W 1, 3/W 2} {
            \node[label] (W \x) [below=of \b]{Worker \x};
            \node[op] (A \x) [right=of W \x] {$A_\x$};
            \node[phantom] (H\x) [right=of A \x] {\phantom{$H(m)$}}
                edge [pre] (A \x);
            \node[op] (B \x) [right=of H\x] {$B_\x$}
                edge [pre] (H\x);
            \node[op] (C \x) [right=of B \x] {$C_\x$}
                edge [pre] (B \x);
        }
        \node[exchange, fit=(H0) (H1) (H2) (H3)] {$\Huge\times$};
        \node[phantom, fit=(W 0) (W 1), inner sep=0pt] (L0) {};
        \node[phantom, fit=(W 2) (W 3), inner sep=0pt] (L2) {};
        \node[process] (P0) [left=of L0] {Process 0};
        \node[process] (P1) [left=of L2] {Process 1};
        \begin{scope}[on background layer]
            \node[rectangle,draw,inner sep=2pt,fit=(P0) (C 0) (C 1)] {};
            \node[rectangle,draw,inner sep=2pt,fit=(P1) (C 2) (C 3)] {};
        \end{scope}
    \end{tikzpicture}
    \caption{Timely dataflow execution model}
    \label{fig:timely_overview}
\end{figure}

A timely dataflow is executed by multiple \emph{workers} (threads) belonging to one or more OS processes, which may reside in one or more machines of a networked cluster.
Workers communicate with each other by exchanging messages over data channels (shared-nothing paradigm) as shown in Figure~\ref{fig:timely_overview}.
Each worker has a local copy of the entire timely dataflow graph and executes all operators in this graph on (disjoint) partitions of the dataflow's input data.
Each worker repeatedly executes dataflow operators concurrent with other workers, sending and receiving data across data exchange channels.
Due to this asynchronous execution model, the presence of concurrent ``in-flight'' timestamps is the rule rather than the exception.

As timely workers execute, they communicate the numbers of logical timestamps they produce and consume to all other workers.
This information allows each worker to determine the possibility that any dataflow operator may yet see any given timestamp in its input.
The timely workers present this information to operators in the form of a \emph{frontier}:

\begin{definition}\label{def:frontier}
A \textbf{frontier} $F$ is a set of logical timestamps such that
\begin{enumerate}[nosep]
\item no element of $F$ is strictly greater than another element of $F$,
\item all timestamps on messages that may still be received are greater than or equal to some element of $F$.
\end{enumerate}
\end{definition}

In many simple settings a frontier is analogous to a \emph{low watermark} in streaming systems like Flink, which indicates the single smallest timestamp that may still be received.
In timely dataflow a frontier must be set-valued rather than a single timestamp because timestamps may be only partially ordered.

Operators in timely dataflow may retain \emph{capabilities} that allow the operator to produce output records with a given timestamp.
All received messages come bearing such a capability for their timestamp.
Each operator can choose to drop capabilities, or downgrade them to later timestamps.
The timely dataflow system tracks capabilities held by operators, and only advances downstream frontiers as these capabilities advance.


Timely dataflow frontiers are the main mechanism for coordination between otherwise asynchronous workers. The frontiers indicate when we can be certain that all messages of a certain timestamp have been received, and it is now safe to take any action that needed to await their arrival. Importantly, frontier information is entirely passive and does not interrupt the system execution; it is up to operators to observe the frontier and determine if there is some work that cannot yet be performed. This enables very fine-grained coordination, without system-level intervention. Further technical details of progress tracking in timely dataflows can be found in \cite{murray:naiad,progress_tracking}.

We will use timely dataflow frontiers to separate migrations into independent arbitrarily
fine-grained timestamps and logically coordinate data movement without using coarse-grained pause-and-resume for parts of the dataflow.

\subsection{Migration formalism and guarantees}\label{sec:mechanism_guarantees}

To frame the mechanism we introduce for live migration in streaming dataflows, we first lay out some formal properties that define correct and live migration.
In the interest of clarity we keep the descriptions casual, but each can be formalized.

We consider stateful dataflow operators that are \emph{data-parallel} and \emph{functional}.
Specifically, an operator acts on input data that are structured as $(\textit{key},\textit{val})$ pairs, each bearing a logical timestamp.
The input is partitioned by its $\textit{key}$ and the operator acts independently on each input partition by sequentially applying each $\textit{val}$ to its state in timestamp order.
For each $\textit{key}$, for each $\textit{val}$ in timestamp order, the operator may change its per-key state arbitrarily, produce arbitrary outputs as a result, and it may schedule further per-key changes at future timestamps (in effect sending itself a new, post-dated $\textit{val}$ for this $\textit{key}$).
\[ \textit{operator}_{\textit{key}}\colon (\textit{state},\, \textit{val}) \rightarrow (\textit{state}',\, [\textit{outputs}],\, [(\textit{vals},\, \textit{times})])\]
The output triples are the new state, the outputs to produce, and future changes that should be presented to the operator.

For a specific \textit{operator}, we can describe the correctness of an implementation. We introduce the notation of \emph{in advance of} as follows.

\begin{definition}[in advance of]\label{def:inadvance} A timestamp \(t\) is \textbf{in advance of}
\begin{enumerate}[nosep]
  \item a timestamp $t'$ if $t$ is greater than or equal to $t'$;
  \item a frontier $F$ if $t$ is greater than or equal to an element of $F$.
\end{enumerate}
\end{definition}

\added{In-advance-of corresponds to the less-or-equal relation for partially ordered sets. For example, a time 6 is in advance of 5.}

\begin{property}[Correctness]\label{prop:correctness}
The \textbf{correct outputs through \textit{time}} are the timestamped outputs that result from each key from the time\-stamp-ordered application of input and post-dated records bearing timestamp not in advance of \textit{time}.
\end{property}

For each migrateable operator, we also consider a \emph{configuration} function, which for each timestamp assigns each key to a specific worker.
\[ \textit{configuration}\colon (\textit{time},\, \textit{key}) \rightarrow \textit{worker} \]
\added{For example, the configuration function could assign a key \texttt{a} to worker 2 for times \([4, 8)\) and to worker 1 for times \([8, 16)\).}

With a specific \textit{configuration}, we can describe the correctness of a migrating implementation.
\begin{property}[Migration]\label{prop:migration}
A computation is \textbf{migrated according to \textit{configuration}} if all updates to \textit{key} with timestamp \textit{time} are performed at worker \textit{configuration(time, key)}.
\end{property}

A configuration function can be represented in many ways, which we will discuss further. In our context we will communicate any changes using a timely dataflow stream, in which configuration changes bear the logical timestamp of their migration. This choice allows us to use timely dataflow's frontier mechanisms to coordinate migrations, and to characterize liveness.

\begin{property}[Completion (liveness)]\label{prop:completion}
  A migrating computation is \textbf{completing} if, once the frontiers of both the data input stream and configuration update stream reach \textit{F}, then (with no further requirements of the input) the output frontier of the computation will eventually reach \textit{F}.
\end{property}

Our goal is to produce a mechanism that satisfies each of these three properties: Correctness, Migration, and Completion.



\subsection{Configuration updates}\label{sec:configuration}

State migration is driven by updates to the \emph{configuration} function introduced in~\ref{sec:mechanism_guarantees}. In \pname these updates are supplied as data along a timely dataflow stream,
each bearing the logical timestamp at which they should take effect. Informally, configuration updates have the form
\[ \textit{update}\colon (\textit{time},\, \textit{key},\, \textit{worker}) \]
indicating that as of \textit{time} the state and values associated with \textit{key} will be located at \textit{worker},
and that this will hold until a new update to \textit{key} is observed with a greater timestamp.
\added{For example, an update could have the form of (\textit{time:} 16, \textit{key:} \texttt{a}, \textit{worker:} 0), which would define the configuration function for times of 16 and beyond.}

As configuration updates are simply data, the user has the ability to drive a migration process by introducing updates as they see fit.
In particular, they have the flexibility to break down a large migration into a sequence of smaller migrations,
each of which have lower duration and between which the system can process data records.
For example, to migrate from one configuration $C_1$ to another $C_2$, a user can use different migration strategies to reveal the changes from $C_1$ to $C_2$:
\begin{description}[nosep]
    \item[All-at-once migration] To simultaneously migrate all keys from $C_1$ to $C_2$, a user could supply all changed (\textit{time}, \textit{key}, \textit{worker}) triples with one common \textit{time}.
        \added{This is essentially an implementation of the \emph{partial pause-and-restart} migration strategy of existing streaming systems as described in Section~\ref{sec:migration-strategies}.}

    \item[Fluid migration] To smoothly migrate keys from $C_1$ to $C_2$, a user could repeatedly choose one key changed from $C_1$ to $C_2$, introduce the new (\textit{time}, \textit{key}, \textit{worker}) triple with the current \textit{time}, and await the migration's completion before choosing the next key.

    \item[Batched migration] To trade off low latency against high throughput, a user can produce batches of changed (\textit{time}, \textit{key}, \textit{worker}) triples with a common \textit{time}, awaiting the completion of the batch before introducing the next batch of changes.
\end{description}
\medskip

We believe that this approach to reconfiguration, as user-supplied data, opens a substantial design space. Not only can users perform fine-grained migration, they can prepare future migrations at specific times, and drive migrations based on timely dataflow computations applied to system measurements. Most users will certainly need assistance in performing effective migration, and we will evaluate several specific instances of the above strategies.

\subsection{\pname's mechanism}\label{sec:mechanism_overview}

\begin{figure}[t]
    \centering
    \begin{subfigure}{\linewidth}
        \centering
        {\includegraphics[width=.8\linewidth,clip,trim=0 1.7cm 0 0]{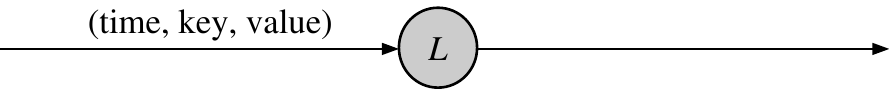}}
        \caption{Original \logicop-operator in a dataflow.}
        \label{fig:mechanism_overview:L}
    \end{subfigure}
    \begin{subfigure}{\linewidth}
        \centering
        {\includegraphics[width=.9\linewidth,trim=0 3pt 0 0pt]{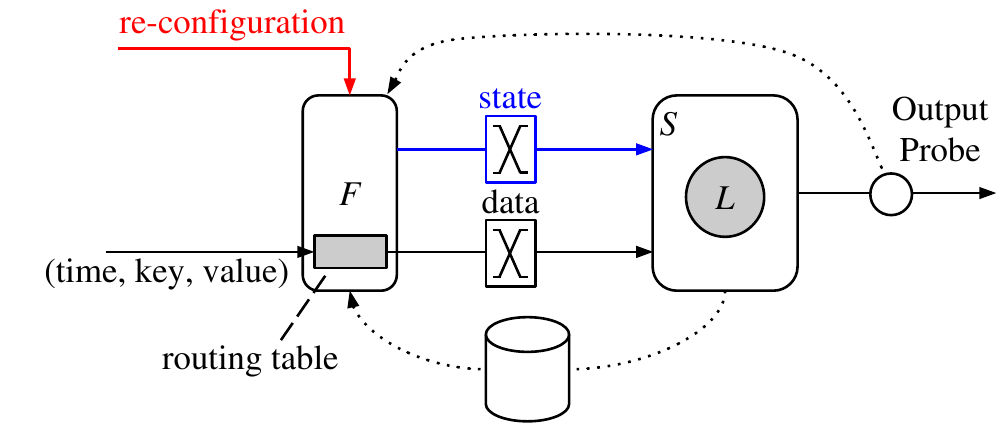}}
        \caption{\pname's operator structure in a dataflow.}
        \label{fig:mechanism_overview:FS}
    \end{subfigure}
    \caption{Overview of \pname's migration mechanism}
    \label{fig:mechanism_overview}
\end{figure}

\begin{figure*}[ht]
    \centering
    \begin{subfigure}[b]{.3\textwidth}
        {\includegraphics[width=.9\linewidth,trim={0 5pt 0 7pt}]{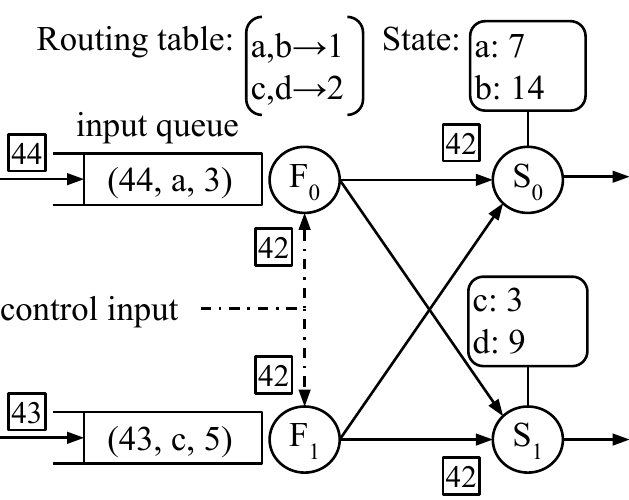}}
        \caption{Before migrating}
        \label{fig:migration_example:before}
    \end{subfigure}
    \begin{subfigure}[b]{.3\textwidth}
        {\includegraphics[width=.9\linewidth,trim={0 5pt 0 7pt}]{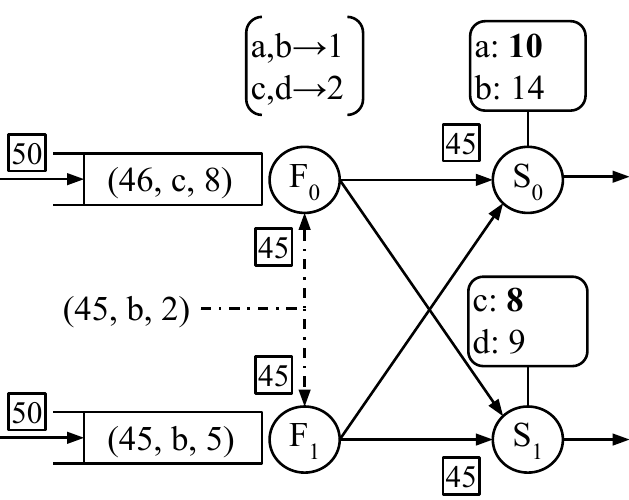}}
        \caption{Receiving a configuration update}
        \label{fig:migration_example:during}
    \end{subfigure}
    \begin{subfigure}[b]{.3\textwidth}
        {\includegraphics[width=.9\linewidth,trim={0 5pt 0 7pt}]{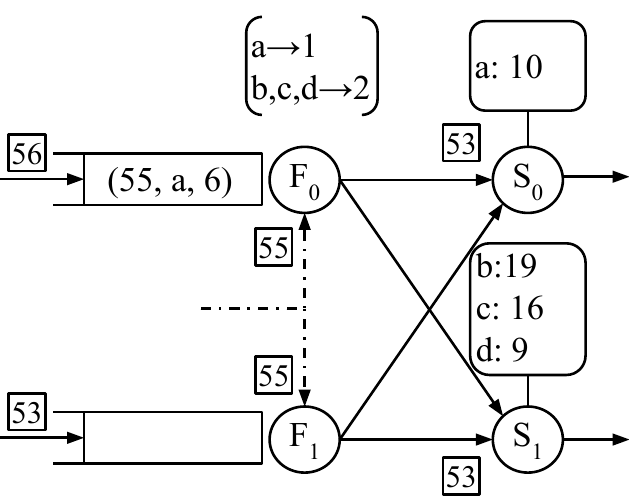}}
        \caption{After migration}
        \label{fig:migration_example:after}
    \end{subfigure}
    \caption{A migrating word-count dataflow executed by two workers. The example is explained in more detail in Section~\ref{sec:example}}
    \label{fig:migration_example}
\end{figure*}

We now describe how to create a migrateable version of an operator \logicop implementing some deterministic, data-parallel \textit{operator} as described in~\ref{sec:mechanism_guarantees}. A non-migrateable implementation would have a single dataflow operator with a single input dataflow stream of (\textit{key}, \textit{val}) pairs, exchanged by \textit{key} before they arrive at the operator.

Instead, we create two operators \fixerop and \stuffop. \fixerop takes the data stream \added{and the stream of configuration updates as an additional input} and produces data pairs and migrating state as outputs. \added{The configuration stream can be ingested from an external controller such as DS2}~\cite{ds2} \added{or Chi}~\cite{Mai18}. \stuffop takes as inputs exchanged data pairs and exchanged migrating state, and applies them to a hosted instance of \logicop, which implements \textit{operator} and maintains both state and pending records for each key.
\Cref{fig:mechanism_overview:FS} presents a schematic overview of the construction.
\added{Recall that in timely dataflow instances of all operators in the dataflow are multiplexed on each worker (core).
The \fixerop and \stuffop on the same worker share access to \logicop's state.}

This construction can be repeated for all the operators in the dataflow that need support for migration. Separate operators can be migrated independently (via separate configuration update streams), or in a coordinated manner by re-using the same configuration update stream. Operators with multiple data inputs can be treated like single-input operators where the migration mechanism acts on both data inputs at the same time.

\stitle{Operator \fixerop}

Operator \fixerop routes (\textit{key}, \textit{val}) pairs according to the configuration at their associated \textit{time}, buffering pairs if \textit{time} is in advance of the frontier of the configuration input. For times in advance of this frontier, the configuration is not yet certain as further configuration updates could still arrive. The configuration at times not in advance of this frontier can no longer be updated. As the data frontier advances, configurations can be retired.

Operator \fixerop is also responsible for initiating state migrations. For a configuration update (\textit{time}, \textit{key}, \textit{worker}), \fixerop must not initiate a migration for \textit{key} until its state has absorbed all updates at times strictly less than \textit{time}. \fixerop initiates a migration once \textit{time} is present in the output frontier of \stuffop, as this implies that there exist no records at timestamps less than \textit{time}, as otherwise they would be present in the frontier in place of \textit{time}.

Operator \fixerop initiates a migration by uninstalling the current state for \textit{key} from its current location in operator \stuffop, and transmitting it bearing timestamp \textit{time} to the instance of operator \stuffop on \textit{worker}. The state includes both the state for \textit{operator}, as well as the list of pending (\textit{val}, \textit{time}) records produced by \textit{operator} for future times.

\stitle{Operator \stuffop}

Operator \stuffop receives exchanged (\textit{key}, \textit{val}) pairs and exchanged state as the result of migrations initiated by \fixerop. \stuffop immediately installs any received state. \stuffop applies received and pending (\textit{key}, \textit{val}) pairs in timestamp order using \textit{operator} once their timestamp is not in advance of either the data or state inputs.

We provide details of \pname's implementation of this mechanism in Section~\ref{sec:implementation}.

\paragraph{Proof sketch}

For each key, $\textit{operator}_\textit{key}$ defines a timeline corresponding to a single-threaded execution, which assigns to each time a pair $(\textit{state},\, [(\textit{val},\, \textit{time})])$ of state and pending records \emph{just before} the application of input records at that time. Let $P(t)$ denote the function from times to these pairs for \textit{key}.

For each key, the \textit{configuration} function partitions this timeline into disjoint intervals, $[t_a, t_b)$, each of which is assigned to one operator instance $\stuffop_a$.

\textbf{Claim}: $\fixerop$ migrates exactly $P(t_a)$ to $\stuffop_a$.

First, $\fixerop$ always routes input records at $time$ to $\stuffop_a$, and so routes all input records in $[t_a, t_b)$ to $\stuffop_a$.
If $\fixerop$ also presents $\stuffop_a$ with $P(t_a)$, it has sufficient input to produce $P(t_b)$. More precisely,

\begin{enumerate}[nosep]
\item Because $\fixerop$ maintains its output frontier at $t_b$, in anticipation of the need to migrate $P(t_b)$, $\stuffop_a$ will apply no input records in advance of $t_b$. And so, it applies exactly the records in $[t_a, t_b)$.
\item Until $\stuffop_a$ transitions to $P(t_b)$, its output frontier will be strictly less than $t_b$, and so $\fixerop$ will not migrate anything other than $P(t_b)$.
\item Because $\fixerop$ maintains its output frontier at $t_b$, and $\stuffop_a$ is able to advance its output frontier to $t_b$, the time $t_b$ will eventually be \emph{in} the output frontier of S.
\end{enumerate}

\subsection{Example}
\label{sec:example}

Figure~\ref{fig:migration_example} presents three snapshots of a migrating streaming word-count dataflow. The figure depicts operator instances $\fixerop_0$ and $\fixerop_1$ of the upstream routing operator, and operator instances $\stuffop_0$ and $\stuffop_1$ of the operator instances hosting the word-count state and update logic. The $\fixerop$ operators maintain input queues of received but not yet routable input data, and an input stream of logically timestamped configuration updates. Although each $\fixerop$ maintains its own routing table, which may temporarily differ from others, we present one for clarity. Input frontiers are represented by boxed numbers, and indicate timestamps that may still arrive on that input.

In Figure~\ref{fig:migration_example:before}, $\fixerop_0$ has enqueued the record (44,
\texttt{a}, 3) and $\fixerop_1$ has enqueued the record (43, \texttt{c}, 5), both because their control input frontier has only reached 42 and so the destination workers at their associated timestamps have not yet been determined. Generally, $\fixerop$ instances will only enqueue records with timestamps in advance of the control input frontier, and the output frontiers of the $\stuffop$ instances can reach the minimum of the data and control input frontiers.

In Figure~\ref{fig:migration_example:during}, both control inputs have progressed
to 45. The buffered records (44, \texttt{a}, 3) and (43,
\texttt{c}, 5) have been forwarded to $\stuffop_1$ and $\stuffop_2$, and the count
operator instances apply the state updates accordingly, shown in bold.
Additionally, both operators have received a configuration update for the key \texttt{b} at time 45. Should the configuration input frontier advance beyond 45, both $\fixerop_0$ and $\fixerop_1$ can integrate the configuration change, and then react.
Operator $\fixerop_0$ would observe that the output frontier of $\stuffop_0$ reaches 45, and initiate a state migration. Operator $\fixerop_1$ would route its buffered input at time 45, to $\stuffop_1$ rather than $\stuffop_0$.

In Figure~\ref{fig:migration_example:after} the migration has completed. Although the configuration frontier has advanced to 55, the output frontiers are held back by the data input frontier of $\fixerop_1$ at 53. \added{According to \cref{def:frontier}, the frontier guarantees that no record with a time earlier than 53 will appear at the input.} If the configuration frontier advances past 55 then operator $\fixerop_0$ could route its queued record, but neither $\stuffop$ operator could apply it until they are certain that there are no other data records that could come before the record at 55.

%

\section{Implementation}\label{sec:implementation}

\pname is an implementation of the migration mechanism described in Section~\ref{sec:mechanism}. In this section, we detail specific choices made in \pname's implementation, including the interfaces used by the application programmer, \pname's specific choices for the grouping and organization of per-key state, and how we implemented \pname's operators in timely dataflow. We conclude with some discussion of how one might implement \pname in other stream processing systems, as well as alternate implementation choices one could consider.



\subsection{\pname's operator interface}\label{sec:interface}

\pname presents users with an operator interface that closely resembles the operator interfaces timely dataflow presents. In several cases, users can use the same operator interface extended only with an additional input stream for configuration updates. More generally, we introduce a new structure to help users isolate and surface all information that must be migrated (state, but also pending future records). These additions are implemented strictly above timely dataflow, but their structure is helpful and they may have value in timely dataflow proper.

The simplest stateful operator interface \pname and timely provide is the \lstinline!state_machine! operator, which takes one input structured as pairs (\textit{key}, \textit{val}) and a state update function which can produce arbitrary output as it changes per-key state in response to keys and values. In \pname, there is an additional input for configuration updates, but the operator signature is otherwise identical.

More generally, timely dataflow supports operators of arbitrary numbers and types of inputs, containing arbitrary user logic, and maintaining arbitrary state. In each case a user must specify a function from input records to integer keys, and the only guarantee timely dataflow provides is that records with the same key are routed to the same worker. Operator execution and state are partitioned by worker, but not necessarily by key.

For \pname to isolate and migrate state and pending work we must encourage users to yield some of the generality timely dataflow provides. However, timely dataflow has already required the user to program partitioned operators, each capable of hosting multiple keys, and we can lean on these idioms to instantiate more fine-grained operators, partitioned not only by worker but further into finer-grained \emph{bins} of keys. Routing functions for each input are already required by timely dataflow, and \pname interposes to allow the function to change according to reconfiguration. Timely dataflow per-worker state is defined implicitly by the state captured by the operator closure, and \pname only makes it more explicit. The use of a helper to enqueue pending work is borrowed from an existing timely dataflow idiom (the \texttt{Notificator}). While \pname's general API is not identical to that of timely dataflow, it is just a more crisp framing of the same idioms.





Listing~\ref{listing:interface} shows how \pname's operator interface is structured.
The interface declares unary and binary stateful operators for single input or dual input operators as well as a state-machine operator.
The logic \added{for the state-machine operator} has to be encoded in the \lstinline!fold!-function. \pname presents data in timestamp order with a corresponding state and notificator object.
Here, migration is transparent and performed without special handling by the operator implementation.


\begin{lstlisting}[float=t,
    breaklines,
    captionpos=b,
    caption={Abstract definition of the \pname operator interfaces. Arguments \texttt{State} and \texttt{Notificator} are provided as mutable references which can be operated upon.},
    label=listing:interface,
    basicstyle=\small\ttfamily,
    escapechar=@,
    ]
fn state_machine(
  control: Stream<ControlInstr>,
  input: Stream<(K, V)>,
  exchange: K -> Integer
  fold: |Key, Val, State| -> List<Output>
) -> Stream<Output>; @\vspace{3pt}@
fn unary(
  control: Stream<ControlInstr>,
  input: Stream<Data>,
  exchange: Data -> Integer,
  fold: |Time, Data, State, Notificator| -> List<Output>
) -> Stream<Output>; @\vspace{3pt}@
fn binary(
  control: Stream<ControlInstr>,
  input1: Stream<Data1>, input2: Stream<Data2>,
  exchange1: Data1 -> Integer,
  exchange2: Data2 -> Integer,
  fold: |Time, Data1, Data2, State, Notificator1, Notificator2| -> List<Output>
) -> Stream<Output>;
\end{lstlisting}

\paragraph{Example}

Listing~\ref{listing:code_word_count} shows an example of a stateful word-count dataflow with a single data input and an additional \texttt{control} input.
The \texttt{stateful\_unary} operator receives the \texttt{control} input, the state type, and a key extraction function as parameters.
The control input carries information about where data is to be routed as discussed in the previous section.
\added{During migration, the state object is converted into a stream of serialized tuples, which are used to reconstruct the object on the receiving worker.}
State is managed in groups of keys, i.e.~many keys of input data will be mapped to the same state object.
The key extraction function defines how this key can be extracted from the input records.

\begin{lstlisting}[float=t,
    breaklines,
    captionpos=b,
    caption={A stateful word-count operator. The operator reads (\emph{word}, \emph{diff})-pairs and outputs the accumulated count of each encountered word. For clarity, the example suppresses details related to Rust's data ownership model.},
    label=listing:code_word_count,
    basicstyle=\small\ttfamily,
    escapechar=@,
    ]
worker.dataflow(|scope| {@\vspace{3pt}@
  // Introduce configuration and input streams.
  let conf = conf_input.to_stream(scope);
  let text = text_input.to_stream(scope);@\vspace{3pt}@
  // Update per-word accumulate counts.
  let count_stream = megaphone::unary(
    conf,
    text,
    |(word, diff)| hash(word),
    |time, data, state, notificator| {
      // map each (word, diff) pair to the accumulation.
      data.map(|(word, diff)| {
        let mut count = state.entry(word).or_insert(0);
        *count += diff;
        (word, count)
      })
    }
  );
});
\end{lstlisting}


\subsection{State organization}


State migration as defined in Section~\ref{sec:mechanism_guarantees} is defined
on a per-key granularity. In a typical streaming dataflow, the number of keys
can be large in the order of million or billions of keys. Managing each key
individually can be costly and thus we selected to group keys into \emph{bins}
and adapt the \textit{configuration} function as follows:
\[\textit{configuration}: (\textit{time}, \textit{bin}) \to \textit{worker}.\]
Additionally, each key is statically assigned to one equivalence class that
identifies the bin it belongs to.

In \pname, the number of bins is configurable in powers of two at startup but
cannot be changed during run-time. A stateful operator gets to see a bin that
holds data for the equivalence class of keys for the current input. Bins are
simply identified by a number, wich corresponds to the most significant bits of
the exchange function specified on the operator.\footnote{Otherwise, keys with similar
least-significant bits are mapped to the same bin; Rust's
\lstinline!HashMap!-implementation suffers from collisions for keys with
similar least-significant bits.}


\pname's mechanism requires two distinct operators, \fixerop and \stuffop. The
operator \stuffop maintains the bins local to a worker and passes references to
the user logic~\logicop. Nevertheless, the \stuffop-operator does not have a
direct channel to its peers. For this reason, \fixerop can obtain a reference
to bins by means of a shared pointer. During a migration, \fixerop serializes
the state obtained via the shared pointer and sends it to the new owning
\stuffop-operator via a regular timely dataflow channel. Note that sharing a
pointer between two operators requires the operators to be executed by the same
process (or thread to avoid synchronization), which is the case for timely
dataflow.

\subsection{Timely instantiation}


In timely dataflow, data is exchanged according to an \emph{exchange} function, which
takes some data and computes an integer representation:
\[ \textit{exchange}: \textit{data} \to \textit{Integer}. \]
Timely dataflow uses this value to decide where to send tuples. In \pname, instead
of assigning data to a worker based on the exchange function, we introduce an
indirection layer where bins are assigned to workers. That way, the exchange
function for the channels from \fixerop to \stuffop is by a specific worker
identifier.



\paragraph{Monitoring output frontiers}

\pname ties migrations to logical time and a computation's progress. A
reconfiguration at a specific time is only to be applied to the system once all
data up to that time has been processed. The $\fixerop$ operators access this information by monitoring the output frontier of the $\stuffop$ operators. Specifically, timely dataflow supports \emph{probes} as a mechanism to observe progress on arbitrary dataflow edges. Each worker attaches a probe to the output stream of the $\stuffop$ operators, and provides the probe to its $\fixerop$ operator instance.

\paragraph{Capturing timely idioms}

For \pname to migrate state, it requires clear isolation of per-key state and pending records. Although timely dataflow operators require users to write operators that can be partitioned across workers, they do not require the state and pending records to be explicitly identified. To simplify programming migrateable operators, we encapsulate several timely dataflow idioms in a helper structure that both manages state and pending records for the user, and surfaces them for migration.

Timely dataflow has a \texttt{Notificator} type that allows an operator to indicate future times at which the operator may produce output, but without encapsulating the keys, states, or records it might use. We implemented an extended notificator that buffers future triples (\textit{time}, \textit{key}, \textit{val}) and can replay subsets for times not in advance of an input frontier. Internally the triples are managed in a priority queue, unlike in timely dataflow, which allows \pname to efficiently maintain large numbers of future triples. By associating data (keys, values) with the times, we relieve the user from maintaining this information on the side. As we will see, \pname's notificator can result in a net reduction in implementation complexity, despite eliciting more information from the user.





\subsection{Discussion}\label{sec:discussion}

Up to now, we explained how to map the abstract model of \pname to an
implementation. The model leaves many details to the implementation, several of
which have a large effect on an implementation's run-time performance. Here, we
want to point out how they interact with other features of the underlying
system, what possible alternatives are and how to integrate \pname into a
larger, controller-based system.

\paragraph{Other systems}

We implemented \pname in timely dataflow, but the mechanisms could be applied on any sufficiently expressive stream processor \added{with support for event time, progress tracking, and state management}. 
Specifically, \added{\pname relies on the ability of} $\fixerop$ operators to 1.\ observe timestamp progress at other locations in the dataflow, 
and 2.\ to extract state from downstream $\stuffop$ operators for migration. 
\added{With regard to first requirement, systems with out-of-band progress tracking like Millwheel}~\cite{Akidau} \added{and Google Dataflow}~\cite{Akidau:2015:DMP:2824032.2824076} \added{also provide the capability to observe dataflow progress externally, while systems with in-band watermaks like Flink would need to provide an additional mechanism.}
\added{Extracting state from downstream operators is straight-forward in timely dataflow where workers manage multiple operators. In systems where each thread of control manages a single operator external coordination and communication mechanisms could be used to effect the same behavior.}

\paragraph{Fault tolerance}


\pname is a library built on timely dataflow abstractions, and inherits fault-tolerance guarantees from the system. 
For example, the Naiad implementation of timely dataflow provides system-wide consistent snapshots, and a \pname implementation on Naiad would inherit fault tolerance.
At the same time, \pname's migration mechanisms effectively provide programmable snapshots on finer granularities, which could feed back into finer-grained fault-tolerance mechanisms.


\paragraph{Alternatives to binning}


\pname's implementation uses binning to reduce the complexity of the
\textit{configuration} function. An alternative to a static mapping of keys to
bins could be achieved by the means of a prefix tree (e.g., a longest-prefix
match as in Internet routing tables).
Extending the functionality of bins to split bins into smaller sets
or merge smaller sets into larger bins would allow run-time reconfiguration of
the actual binning strategy rather than setting it up during initialization
without the option to change it later on.

\paragraph{Migration controller}

We implemented \pname as a system that provides an input for configuration updates
to be supplied by an external controller. \added{The only requirement \pname places on the controller is to adhere to the control command format as described in Section~\ref{sec:configuration}.} A controller could observe the performance characteristics of a computation on a per-key level and correlate this with the input workload. For example, the recent DS2~\cite{ds2} system automatically measures and re-scales streaming systems to meet throughput targets. \added{\pname can also be driven by general re-configuration controllers and is not restricted to elasticity policies. For instance, the configuration stream could also be provided by Dhalion}~\cite{Floratou2017} \added{or Chi}~\cite{Mai18}.

Independently, we have observed and implemented several details for effective migration. Specifically, we can use bipartite matching to group migrations that do not interfere with each other, reducing the number of migration steps without much increasing the maximum latency. We can also insert a gap between migrations to allow the system to immediately drain enqueued records, rather than during the next migration, which reduces the maximum latency from two migration durations to just one.

\section{Evaluation}\label{sec:evaluation}

Our evaluation of \sys is in three parts.  We are interested in
particular in the latency of streaming queries, and how they are
affected by \sys both in a steady state (where no migration is
occuring) and during a migration
operation.

First, in Section~\ref{sec:nexmark} we use the NEXMark benchmarking
suite~\cite{nexmark_queries,nexmark_paper} to \added{compare
\sys with prior techniques} under a realistic workload. NEXMark consists of
queries covering a variety of operators and windowing behaviors.
Next, in Section~\ref{sec:eval:overhead} we look at the overhead of \sys when no migration occurs: this is
the cost of providing migration functionality in stateful dataflow
operators, versus using optimized operators which cannot migrate
state.
Finally, in Section~\ref{sec:performance} we use a microbenchmark to
investigate how parameters like the number of bins and size of the
state affect migration performance.

We run all experiments on a cluster of four machines, each with four
Intel Xeon E5-4650~v2 @\SI{2.40}{\giga\hertz} CPUs (each 10 cores with
hyperthreading) and \SI{512}{\gibi\byte} of RAM, running Ubuntu
\texttt{18.04}. For each experiment, we pin a timely process with four
workers to a single CPU socket.
Our open-loop testing harness supplies the input at a specified rate, even if the
system itself becomes less responsive (e.g., during a migration). We
record the observed latency every \SI{250}{\milli\second}, in units of
nanoseconds, which are recorded in a histogram of
logarithmically-sized bins.

Unless otherwise specified, we migrate the state of the main operator of each dataflow.
We initially migrate half of the keys on half of the workers to the other half of the workers (25\% of the total state), which results in an imbalanced assignment. We then perform and report the details of a second migration back to the balanced configuration.


\newcommand{\figwi}{.88\linewidth}

\newcommand{\querytitle}[1]{\textbf{#1}}

\subsection{NEXMark Benchmark}\label{sec:nexmark}

The NEXMark suite models an auction site in
which a high-volume stream of users, auctions, and bids arrive, and
eight standing queries are maintained reflecting a variety of
relational queries including stateless streaming
transformations (e.g., map and filter in Q1 and Q2
respectively), a stateful record-at-a-time two-input operator
(incremental join in Q3), and various window operators (e.g.,
sliding window in Q5, tumbling window join in Q8), and
complex multi-operator dataflows with shared components (Q4 and
Q6).

We have implemented all eight of the NEXMark queries in both native
timely dataflow and using \pname.
Table~\ref{tab:nexmark_impl_complexity} lists the lines of code
for queries 1--8. \emph{Native} is a hand-tuned implementation,
\emph{\pname} is implemented using the stateful operator
interface. Note that the implementation complexity for the native
implementation is higher in most cases as we include optimizations
from Section~\ref{sec:implementation} which are not offered by the
system but need to be implemented for each operator by hand.

\begin{table}[t]
    \centering
    \caption{NEXMark query implementations lines of code.}
    \label{tab:nexmark_impl_complexity}
    \begin{tabular}{lrrrrrrrr}
                & Q1 & Q2 & Q3 & Q4 & Q5 & Q6 & Q7 & Q8 \tabularnewline
        \midrule
        Native  & 12 & 14 & 58 & 128& 73 & 130& 55 & 58 \tabularnewline
        \pname  & 16 & 18 & 41 & 74 & 46 & 74 & 54 & 29 \tabularnewline
    \end{tabular}
\end{table}

To test our hypothesis that \sys supports efficient migration on realistic workloads, we run each NEXMark query under high load and migrate the state of each query without interrupting the query processing itself.
Our test harness uses a reference input data generator and increases its rate.
The data generator can be played at a higher rate but this does not change
certain intrinsic properties.
For example, the number of active auctions is static, and so increasing the event rate decreases auction duration.
For this reason, we present time-dilated variants of queries Q5 and Q8 containing large time-based windows (up to 12 hours).
We run all queries with \num{4e6} updates per second.
For stateful queries, we perform a first migration at \SI{400}{\second} and perform and report a second re-balancing migration at \SI{800}{\second}.
We compare \added{\emph{all-at-once}, which is essentially equivalent to the partial pause-and-restart strategy adopted by existing systems, and \emph{batched}, \pname's optimized migration strategy which strikes a balance between migration latency and duration (cf. Section~\ref{sec:configuration}).} We use $2^{12}$ bins for \pname's migration; in Section~\ref{sec:eval:overhead} we study \pname's sensitivity to the bin count.

\begin{figure}[p]
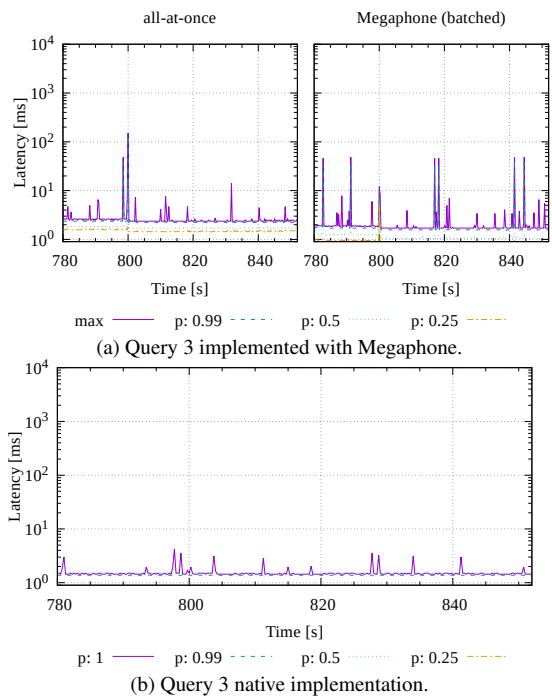

    \centering
    \begin{subfigure}{\linewidth}
        \centering
        {\includegraphics[width=\figwi,page=3,trim={0 6pt 0 14pt}]{figures/charts/c27227cdd6648c84/plot_latency_timeline+bin_shift=12+binary=timely+duration=1200+fake_stateful=False+final_config=uniform_skew+initial_config=uniform+machine_local=False+processes=4+queries=q3-flex+rate=4000000+time_dilation=1+workers=4.pdf}}
        \caption{Query 3 implemented with \pname.}
        \label{fig:nx_q3_timeline:megaphone}
    \end{subfigure}

    \begin{subfigure}{\linewidth}
        \centering
        {\includegraphics[width=\figwi,page=3,trim={0 6pt 0 20pt}]{figures/charts/c27227cdd6648c84/plot_latency_timeline+bin_shift=12+binary=timely+duration=1200+fake_stateful=False+final_config=uniform_skew+initial_config=uniform+machine_local=False+processes=4+queries=q3+rate=4000000+time_dilation=1+workers=4.pdf}}
        \caption{Query 3 native implementation.}
        \label{fig:nx_q3_timeline:native}
    \end{subfigure}

    \caption{NEXMark query latency for Q3. A small latency spike can be observed at \SI{800}{\second} for both all-at-once and batched migration strategies, reaching more than \SI{100}{\milli\second} for all-at-once and \SI{10}{\milli\second} for batched migration. Although the state for query 3 grows without bounds, this did not bear significance after \SI{800}{\second}.}
    \label{fig:nx_q3_timeline}
\end{figure}

\begin{figure}[p]
    \centering

    {\includegraphics[width=\figwi,page=3,page=3,trim={0 6pt 0 12pt}]{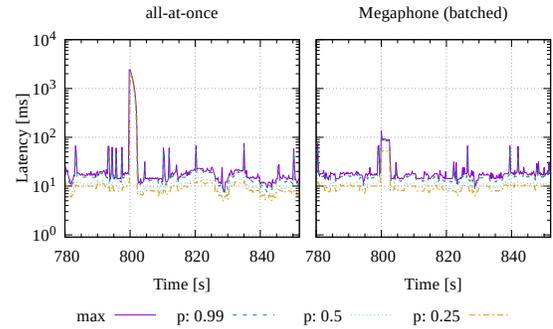}}

    \caption{NEXMark query latency for Q4, \num{4e6} requests per second, reconfiguration at \SI{800}{\second}.}

    \label{fig:nx_q4_timeline}
\end{figure}

\begin{figure}[p]
    \centering

    {\includegraphics[width=\figwi,page=3,trim={0 6pt 0 12pt}]{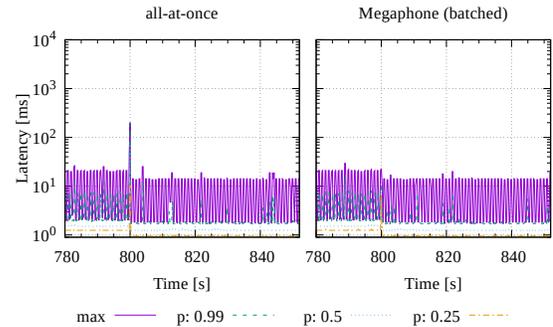}}

    \caption{NEXMark query latency for Q5, \num{4e6} requests per second, reconfiguration at \SI{800}{\second} with time dilation.}

    \label{fig:nx_q5_timeline}
\end{figure}

\begin{figure}[p]
    \centering
    {\includegraphics[width=\figwi,page=3,trim={0 6pt 0 12pt}]{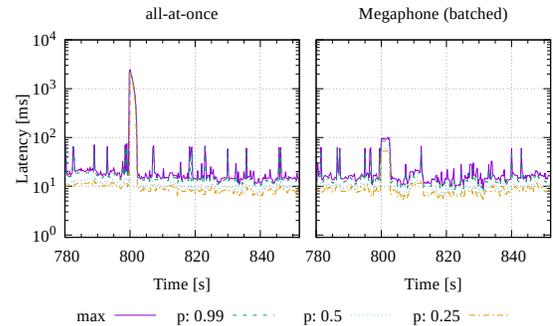}}
    \caption{NEXMark query latency for Q6, \num{4e6} requests per second, reconfiguration at \SI{800}{\second}.}
    \label{fig:nx_q6_timeline}
\end{figure}

\begin{figure}[p]
    \centering

    {\includegraphics[width=\figwi,page=3,trim={0 6pt 0 12pt}]{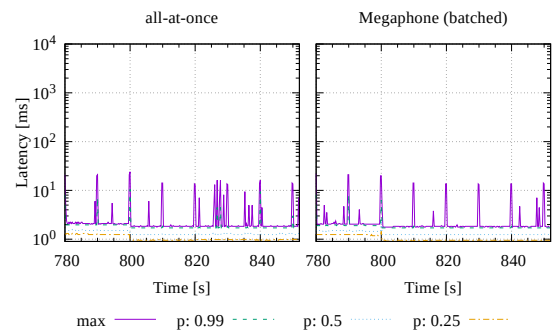}}

    \caption{NEXMark query latency for Q7, \num{4e6} requests per second, reconfiguration at \SI{800}{\second}.}

    \label{fig:nx_q7_timeline}
\end{figure}

\begin{figure}[htb]
    \centering

    {\includegraphics[width=\figwi,page=3,trim={0 6pt 0 12pt}]{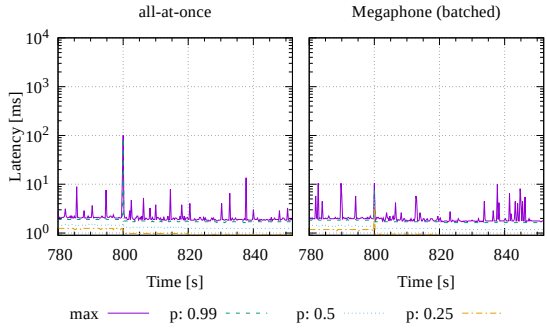}}

    \caption{NEXMark query latency for Q8, \num{4e6} requests per second, reconfiguration at \SI{800}{\second} with time dilation.}

    \label{fig:nx_q8_timeline}
\end{figure}

Figures~\ref{fig:nx_q3_timeline} through \ref{fig:nx_q8_timeline} show timelines for the second migration of stateful queries Q3 through Q8. Generally, the all-at-once migrations experience maximum latencies proportional to the amount of state maintained, whereas the latencies of \pname's batched migration are substantially lower when the amount of state is large.

\begin{figure}[p]
    \centering

    {\includegraphics[width=\figwi,page=1,trim={0 6pt 0 28pt}]{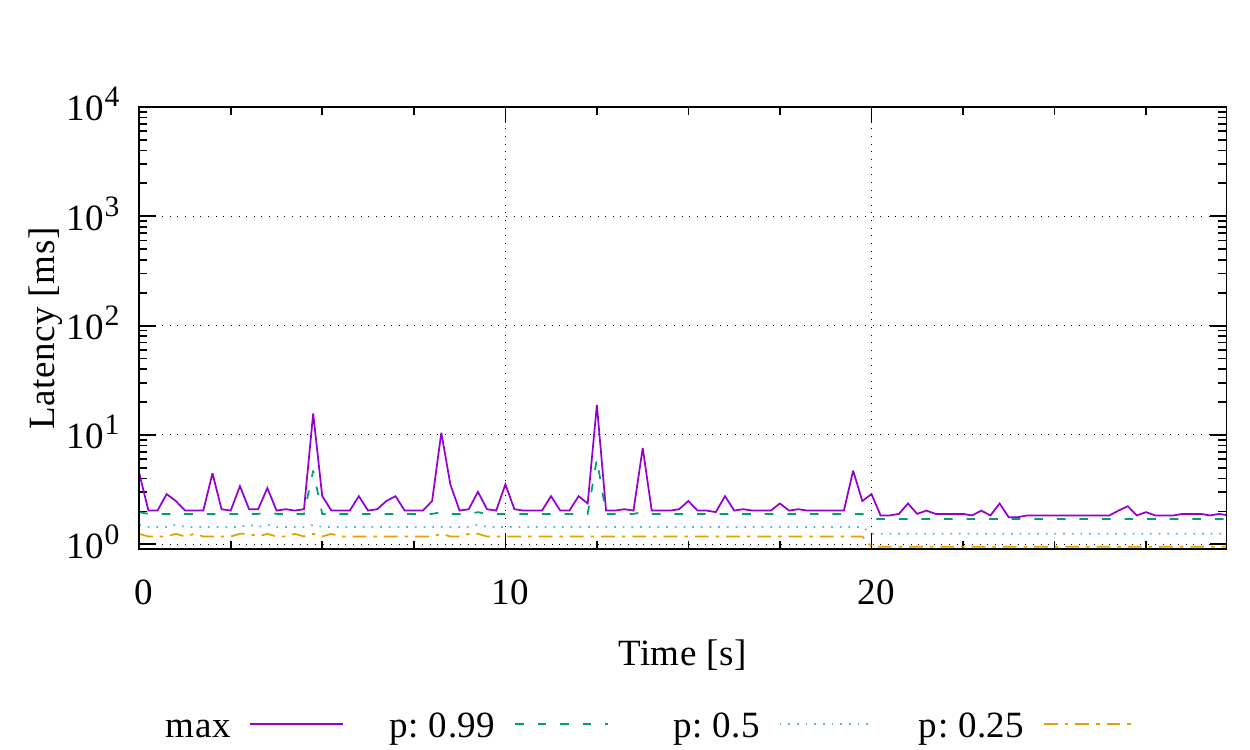}}

    \caption{NEXMark query latency for Q1, \num{4e6} requests per second, reconfiguration at \SI{10}{\second} and \SI{20}{\second}. No latency spike occurs during migration as the query does not accumulate state.}

    \label{fig:nx_q1_timeline}
\end{figure}
\begin{figure}[p]
    \centering

    {\includegraphics[width=\figwi,page=1,trim={0 6pt 0 29pt}]{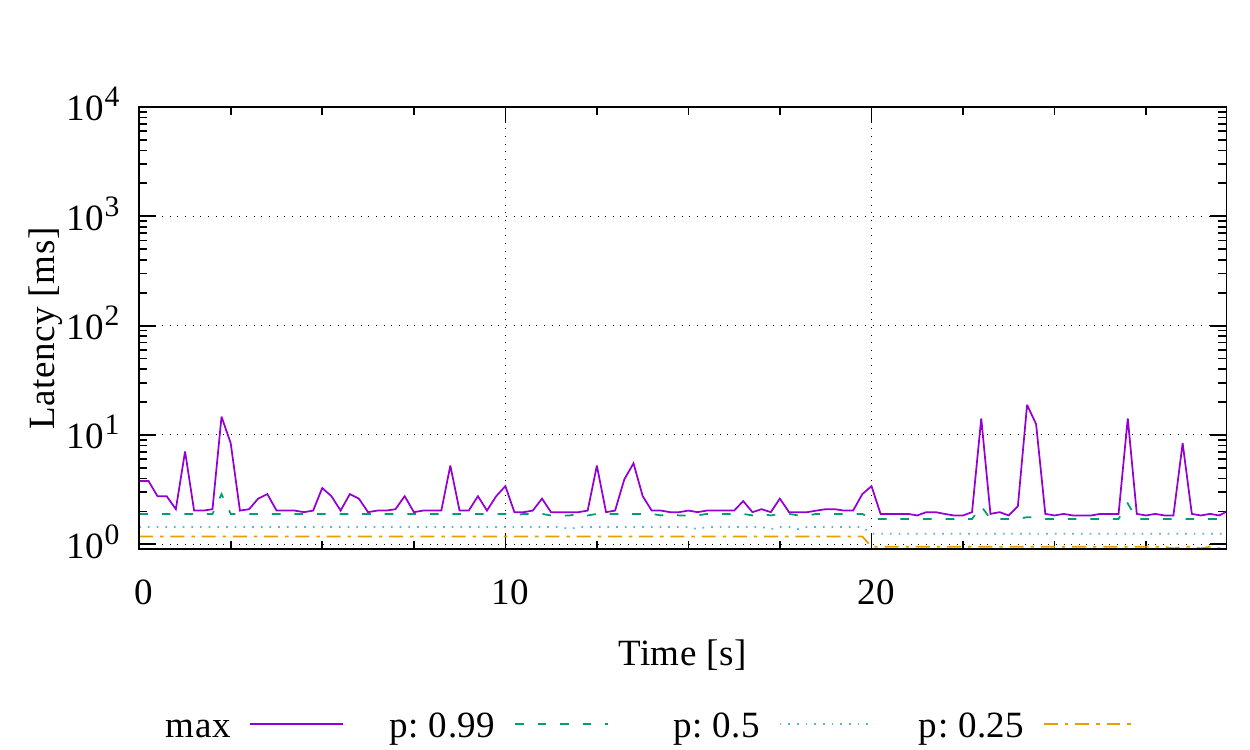}}

    \caption{NEXMark query latency for Q2, \num{4e6} requests per second, reconfiguration at
    \SI{10}{\second} and \SI{20}{\second}. No latency spike occurs during migration as the query does not accumulate state.}

    \label{fig:nx_q2_timeline}
\end{figure}

\querytitle{Query 1 and Query 2} maintain no state.
Q1 transforms the stream of bids to use a different currency, while Q2 filters bids by their auction identifiers.
Despite the fact that both queries do not accumulate state to migrate, we demonstrate their behavior to establish a baseline for \pname and our test harness.
Figures~\ref{fig:nx_q1_timeline} and~\ref{fig:nx_q2_timeline} show query latency during two migrations where no state is thus transferred; any impact is dominated by system noise.

\querytitle{Query 3} joins auctions and people to recommend local auctions to individuals.
The join operator maintains the auctions and people relations, using the seller and person as the keys, respectively.
This state grows without bound as the computation runs.
Figure~\ref{fig:nx_q3_timeline} shows the query latency for both \sys, and
the native timely implementation. We note that while the native timely implementation has some spikes, they are more pronounced in \pname, whose tail latency we investigate further in Section~\ref{sec:eval:overhead}.


\querytitle{Query 4} reports the average closing prices of auctions in
a category relying on a stream of closed auctions, derived
from the streams of bids and auctions, which we compute and
maintain, and contains one operator keyed by auction id
which accumulates relevant bids until the auction closes, at which
point the auction is reported and removed. The NEXMark generator is
designed to have a fixed number of auctions at a time, and so the
state remains bounded. Figure~\ref{fig:nx_q4_timeline} shows the latency timeline during the second migration.
The all-at-once migration strategy causes a latency spike of more than two seconds whereas the batched migration strategy only shows an increase in latency of up to \SI{100}{\milli\second}.

\querytitle{Query 5} reports, each minute, the auctions with the highest
number of bids taken over the previous sixty minutes.  It maintains up
to sixty counts for each auction, so that it can both report and
retract counts as time advances. To elicit more regular behavior, our implementation reports every second over the previous minute, effectively dilating time by a factor of 60.
Figure~\ref{fig:nx_q5_timeline} shows the latency timeline for the second migration; the all-at-once migration is an order of magnitude larger than the per-second events, whereas \pname's batched migration is not distinguishable.

\querytitle{Query 6} reports the average closing price for the
last ten auctions of each seller. This operator is keyed by auction
seller, and maintains a list of up to ten prices. As the computation
proceeds, the set of sellers, and so the associated state, grows
without bound. Figure~\ref{fig:nx_q6_timeline} shows the timeline at the second migration.
The result is similar to query 4 because both have a large fraction of the query plan in common.

\querytitle{Query 7} reports the highest bid each minute, and the
results are shown in Figure~\ref{fig:nx_q7_timeline}. This query
has minimal state (one value) but does require a data exchange to
collect worker-local aggregations to produce a computation-wide
aggregate.
Because the state is so small, there is no distinction between all-at-once and \pname's batched migration.

\querytitle{Query 8} reports a twelve-hour windowed join between new
people and new auction sellers. This query has the potential to
maintain a massive amount of state, as twelve hours of auction and
people data is substantial. Once reached, the peak size of state is
maintained. To show the effect of twelve-hour windows, we dilate the internal time by a factor of 79. The reconfiguration time of \SI{800}{\second} corresponds to approximately \SI{17.5}{\hour} of event time.


These results show that for NEXMark queries maintaining large amounts of state, all-at-once migration can introduce significant disruption, which \pname's batched migration can mitigate. In principle, the latency could be reduced still further with the fluid migration strategy, which we evaluate in Section~\ref{sec:performance}.




\subsection{Overhead of the interface}\label{sec:eval:overhead}

We now use a counting microbenchmark to measure the overhead of \pname, from which one can determine an appropriate trade-off between migration granularity and this overhead.
We compare \pname to native timely dataflow implementations, as we vary the number of bins that \pname uses for state. We anticipate that this overhead will increase with the number of bins, as \pname must consult a larger routing table.

The workload uses a stream of randomly selected 64-bit integer identifiers, drawn uniformly from a domain defined per experiment.
The query reports the cumulative counts of the number of times each identifier has occurred.
In these workloads, the state is the per-identifier count, intentionally small and simple so that we can see the effect of migration rather than associated computation. We consider two variants, an implementation that uses hashmaps for bins (``hash count''), and an optimized implementation that uses dense arrays to remove hashmap computation (``key count'').

Each experiment is parameterized by a domain size (the number of distinct keys) and an input rate (in records per second), for which we then vary the number of bins used by \pname.
Each experiment pre-loads one instance of each key to avoid measuring latency due to state re-allocation at
runtime.

\begin{figure}[tb]
    \centering
    \begin{subfigure}{1\linewidth}
        \centering
        {\includegraphics[width=\figwi,trim={0 6pt 0 12pt}]{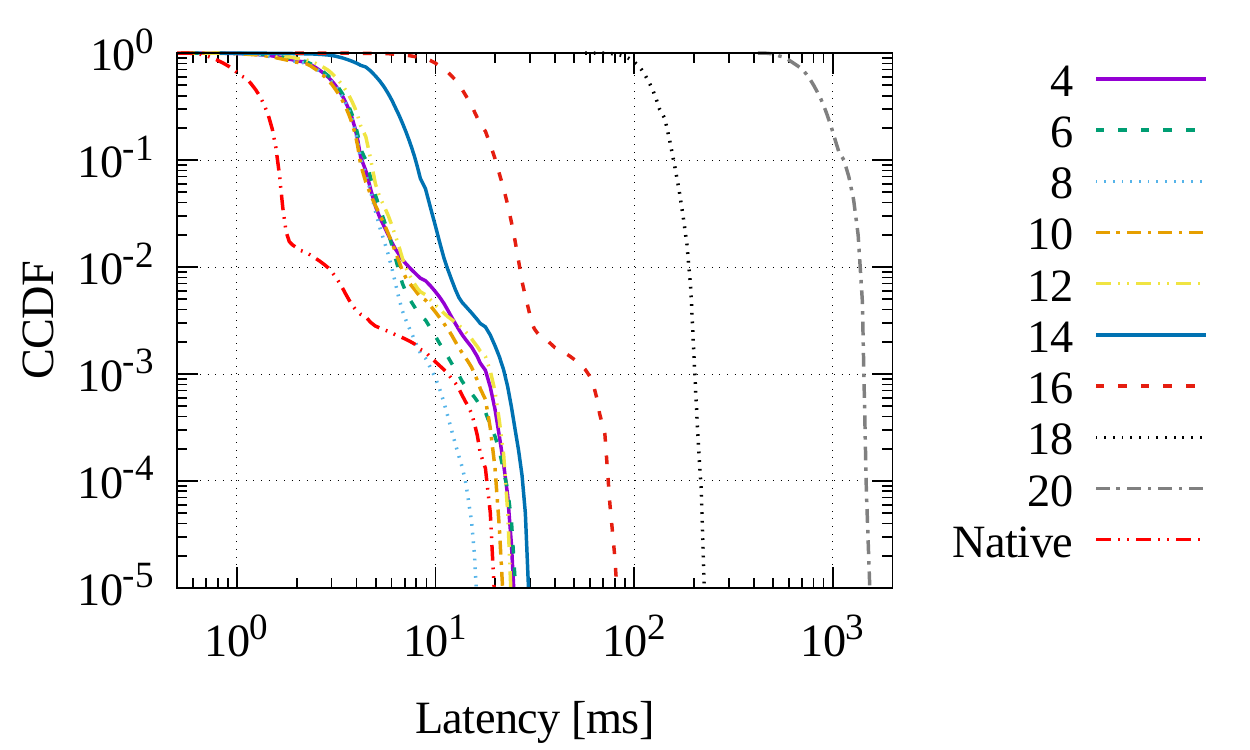}}
        \subcaption{\glsentrytext{ccdf} of per-record latencies}
        \label{fig:wc_baseline_hash_256:ccdf}
    \end{subfigure}

    \begin{subtable}{1\linewidth}
        \centering\small
        \begin{tabular}{lrrrr}
            \input{figures/charts/c27227cdd6648c84/plot_migration_queries_latency_hashmap+backend=hashmapnative+binary=word_count+domain=256000000+duration=30+fake_stateful=False+final_config=uniform+initial_config=uniform+machine_local=False+migration=sudden+processes=4+rate=4000000+workers=4_table.tex}
        \end{tabular}
        \subcaption{Selected percentiles and their latency in \si{\milli\second}}
        \label{fig:wc_baseline_hash_256:tab}
    \end{subtable}

    \caption{Hash-count overhead experiment with
    \num{256e6} unique keys and an update rate of \num{4e6} per second.
    Experiment numbers in~\subref{fig:wc_baseline_hash_256:ccdf}
    and~\subref{fig:wc_baseline_hash_256:tab} indicate \(\log\) bin count.}

    \label{fig:wc_baseline_hash_256}
\end{figure}

\begin{figure}[ht]
    \centering
    \begin{subfigure}{1\linewidth}
        \centering
        {\includegraphics[width=\figwi,trim={0 6pt 0 12pt}]{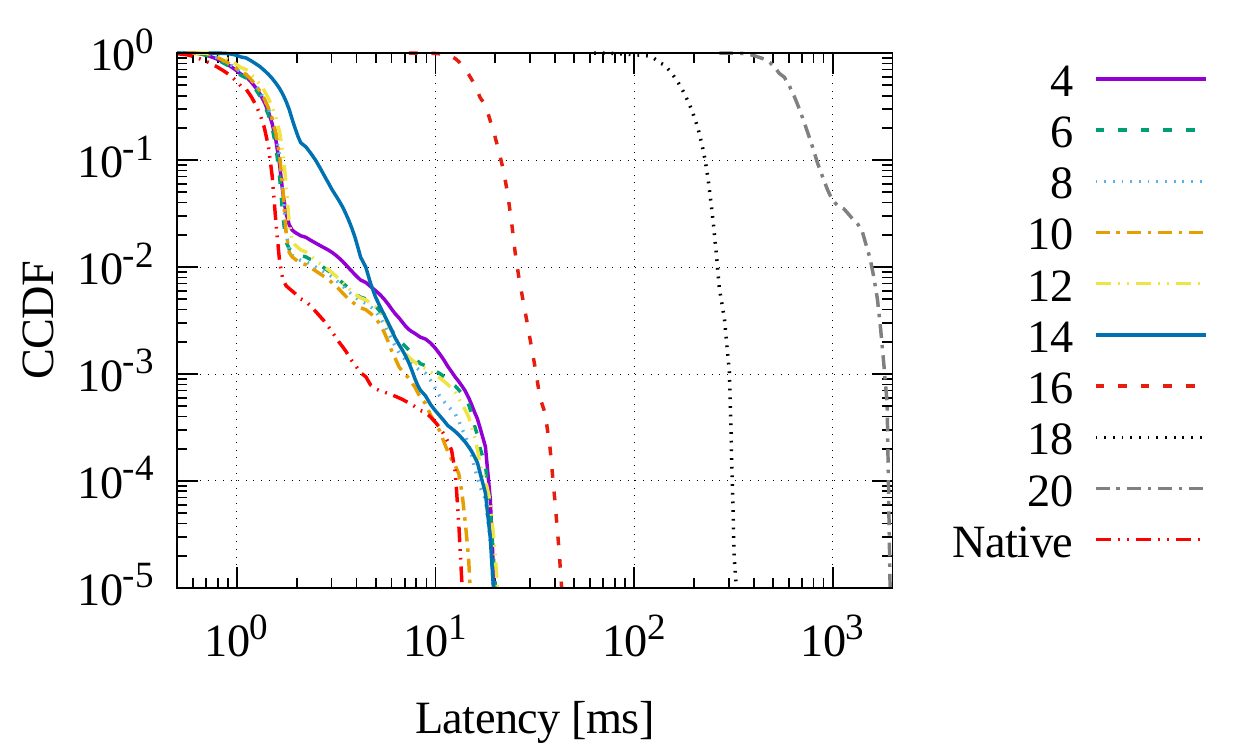}}
        \subcaption{\glsentrytext{ccdf} of per-record latencies}
        \label{fig:wc_baseline_vec_256:ccdf}
    \end{subfigure}

    \begin{subtable}{1\linewidth}
        \centering\small
        \begin{tabular}{lrrrr}
            \input{figures/charts/c27227cdd6648c84/plot_migration_queries_latency_vec+backend=vecnative+binary=word_count+domain=256000000+duration=30+fake_stateful=False+final_config=uniform+initial_config=uniform+machine_local=False+migration=sudden+processes=4+rate=4000000+workers=4_table.tex}
        \end{tabular}
        \subcaption{Selected percentiles and their latency in \si{\milli\second}}
        \label{fig:wc_baseline_vec_256:tab}
    \end{subtable}

    \caption{Key-count overhead experiment with
    \num{256e6} unique keys and an update rate of \num{4e6} per second.
    Experiment numbers in~\subref{fig:wc_baseline_vec_256:ccdf}
    and~\subref{fig:wc_baseline_vec_256:tab} indicate \(\log\) bin count.}

    \label{fig:wc_baseline_vec_256}
\end{figure}

\begin{figure}[ht]
    \centering
    \begin{subfigure}{1\linewidth}
        \centering
        {\includegraphics[width=\figwi,trim={0 6pt 0 14pt}]{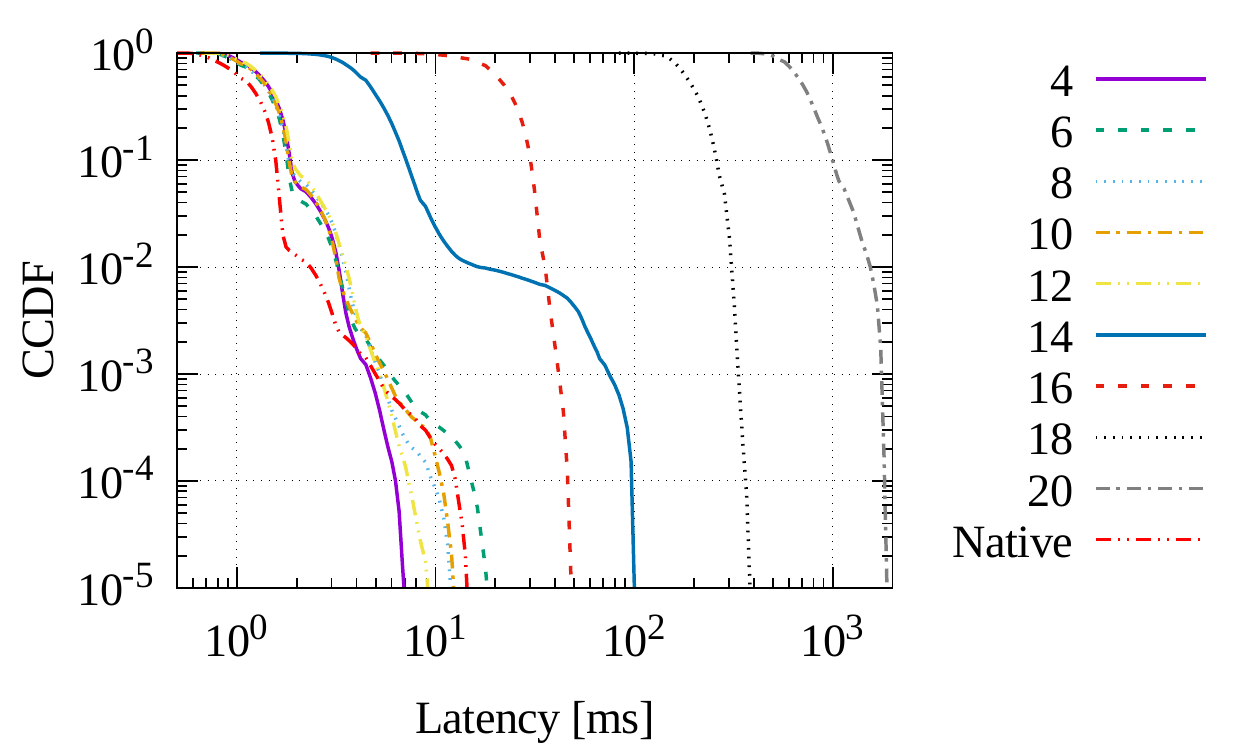}}
        \subcaption{\glsentrytext{ccdf} of per-record latencies}
        \label{fig:wc_baseline_vec_8192:ccdf}
    \end{subfigure}

    \begin{subtable}{1\linewidth}
        \centering\small
        \begin{tabular}{lrrrr}
            \input{figures/charts/c27227cdd6648c84/plot_migration_queries_latency_vec+backend=vecnative+binary=word_count+domain=8192000000+duration=30+fake_stateful=False+final_config=uniform+initial_config=uniform+machine_local=False+migration=sudden+processes=4+rate=4000000+workers=4_table.tex}
        \end{tabular}
        \subcaption{Selected percentiles and their latency in \si{\milli\second}}
        \label{fig:wc_baseline_vec_8192:tab}
    \end{subtable}

    \caption{Key-count overhead experiment with
    \num{8192e6} unique keys and an update rate of \num{4e6} per second.
    Experiment numbers in~\subref{fig:wc_baseline_vec_8192:ccdf}
    and~\subref{fig:wc_baseline_vec_8192:tab} indicate \(\log\) bin count.}

    \label{fig:wc_baseline_vec_8192}
\end{figure}

Figure~\ref{fig:wc_baseline_hash_256} shows the \gls{ccdf} of per-record latency for the hash-count experiment with \num{256e6} distinct keys and a rate of \num{4e6} updates per second.
Figure~\ref{fig:wc_baseline_vec_256} shows the \gls{ccdf} of per-record latency for the key-count experiment with \num{256e6} distinct keys and a rate of \num{4e6} updates per second.
Figure~\ref{fig:wc_baseline_vec_8192} shows the \gls{ccdf} of per-record latency for the key-count experiment with \num{8192e6} distinct keys and a rate of \num{4e6} updates per second.
Each figure reports measurements for a native timely dataflow implementation, and for \pname with geometrically increasing numbers of bins.

For small bin counts, the latencies remain a small constant factor larger than the native implementation, but this increases noticeably once we reach $2^{16}$ bins.
We conclude that while a performance penalty exists, it can be an acceptable trade-off for
flexible stateful dataflow reconfiguration. A bin-count parameter of up to \(2^{12}\) leads to largely indistinguishable results, and we will use this number when we need to hold the bin count constant in the rest of the evaluation.

\subsection{Migration micro-benchmarks}\label{sec:performance}

We now use the counting benchmark from the previous section to analyse how
various parameters influence the maximum latency and duration of \pname during a migration.
Specifically,
\begin{enumerate}[nosep]
    \item In Section~\ref{sec:eval:bins} we evaluate the maximum latency and duration of migration strategies \textbf{as the number of bins increases}. We expect \pname's maximum latencies to decrease with more bins, without affecting duration.
    \item In Section~\ref{sec:eval:keys} we evaluate the maximum latency and duration of migration strategies \textbf{as the number of distinct keys increases}. We expect all maximum latencies and durations to increase linearly with the amount of maintained state.
    \item In Section~\ref{sec:eval:migr_const} we evaluate the maximum latency and duration of migration strategies \textbf{as the number of distinct keys and bins increase proportionally}. We expect that with a constant per-bin state size \pname will maintain a fixed maximum latency while the duration increases.
    \item \added{In Section~\ref{sec:eval:tp} we evaluate \textbf{the latency under load} during migration and steady-state. We expect a smaller maximum latency for \pname migrations.}
    \item In Section~\ref{sec:eval:mem} we evaluate \textbf{the memory consumption} during migration. We expect a smaller memory footprint for \pname migrations.
\end{enumerate}

Each of our migration experiments largely resembles the shapes seen in Figure~\ref{fig:wc_migration_timeline}, where each migration strategy has a well defined \emph{duration} and \emph{maximum latency}. For example, the all-at-once migration strategy has a relatively short duration with a large maximum latency, whereas the bin-at-a-time (\emph{fluid}) migration strategy has a longer duration and lower maximum latency, and the batched migration strategy lies between the two. In these experiments we summarize each migration by the duration of the migration, and the maximum latency observed during the migration.

\subsubsection{Number of bins vary}\label{sec:eval:bins}

We now evaluate the behavior of different migration strategies for varying numbers of bins. As we increase the number of bins we expect to see fluid and batched migration achieve lower maximum latencies, though ideally with relatively unchanged durations. We do not expect to see all-at-once migration behave differently as a function of the number of bins, as it conducts all of its migrations simultaneously.

Holding the rates and bin counts fixed, we will vary the number of bins from $2^4$ up to $2^{14}$ by factors of four. For each configuration, we run for one minute to establish a steady state, and then initiate a migration and continue for one another minute. During this whole time the rate of input records continues uninterrupted.

\begin{figure}[ht]
    \centering

    {\includegraphics[width=\figwi,page=2,trim={0 6pt 0 16pt}]{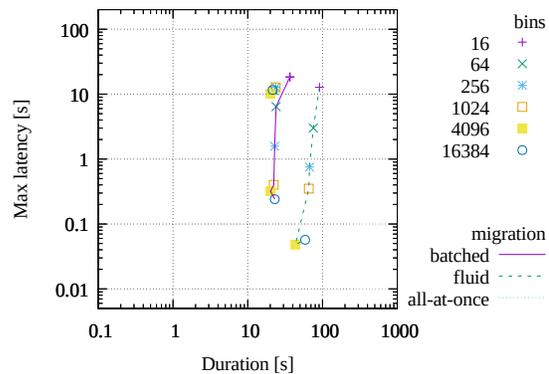}}

    \caption{Key-count migration latency vs.~duration, varying bin
    count for a fixed domain of \num{4096e6} keys. The vertical lines indicate that increasing the granularity of migration can reduce maximum latency for fluid and batched migrations without increasing the duration. The all-at-once migration datapoints remain in a cluster independent of the migration granularity.}

    \label{fig:wc_migration_breakdown_bin_shift}
\end{figure}

Figure~\ref{fig:wc_migration_breakdown_bin_shift} reports the latency-vs-duration trade-off of the three migration strategies as we vary the number of bins. The connected lines each describe one strategy, and the common shapes describe a common number of bins. We see that all all-at-once migration experiments are in a low duration high latency cluster. Both fluid and batched migration achieve lower maximum latency as we increase the number of bins, without negatively impacting the duration.

\subsubsection{Number of keys vary}\label{sec:eval:keys}

We now evaluate the behavior of different migration strategies for varying domain sizes. Holding the rates and bin counts fixed, we will vary the number of keys from \num{256e6} up to \num{8192e6} by factors of two. For each configuration, we run for one minute to establish a steady state, and then initiate a migration and continue for one another minute. During this whole time the rate of input records continues uninterrupted.

\begin{figure}[tb]
    \centering
    {\includegraphics[width=\figwi,page=2,trim={0 6pt 0 16pt}]{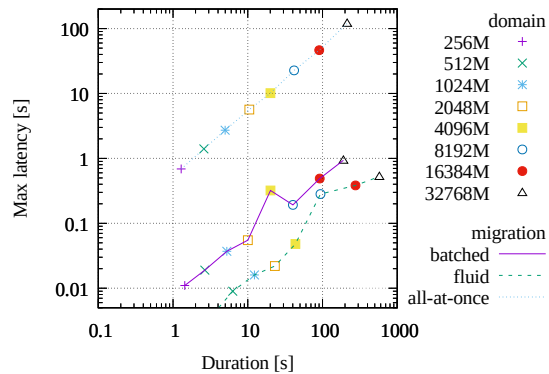}}

    \caption{Key-count migration latency vs.~duration, varying domain
    for a fixed rate of \num{4e6}. As the domain size increases the migration granularity increases, and the duration and maximum latencies increase proportionally.}

    \label{fig:wc_migration_breakdown_domain}
\end{figure}

Figure~\ref{fig:wc_migration_breakdown_domain} reports the latency-vs-duration trade-off of the three migration strategies as we vary the number of distinct keys. The connected lines each describe one strategy, and the common shapes describe a common number of distinct keys. We see that for any experiment, all-at-once migration has the highest latency and lowest duration, fluid migration has a lower latency and higher duration, and batched migration often has the best qualities of both.

\subsubsection{Number of keys and bins vary proportionally}
\label{sec:eval:migr_const}

In the previous experiments, we either fixed the number of bins or the number of keys
while varying the other parameter. In this experiment, we vary both bins and keys together such that the total amount of data per bin stays constant.
This maintains a fixed migration granularity, which should have a fixed maximum latency even as the number of keys (and total state) increases.
We run the key count experiment and fix the number
of keys per bin to \num{4e6}. We then increase the domain in steps of powers
of two starting at \num{256e6} and increase the number of bins such that the
keys per bin stays constant. The maximum domain is \num{32e9} keys.

\begin{figure}[t]
    \centering
    {\includegraphics[width=\figwi,page=2,trim={0 6pt 0 16pt}]{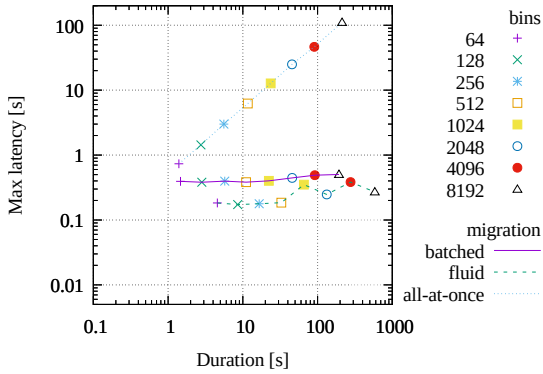}}
    \caption{Latency and duration of key-count migrations for fixed state per bin. By holding the granularity of migration fixed, the maximum latencies of fluid and batched migration remain fixed even as the durations of all strategies increase.}
    \label{fig:wc_migration_breakdown_constant_bin}
\end{figure}

Figure~\ref{fig:wc_migration_breakdown_constant_bin} reports the
latency-versus-duration trade-off for the three migration strategies as we
increase domain and number of bins while keeping the state per bin constant.
The lines describe one migration strategy and the points describe a different
configuration. We can observe that for fluid and batched migration the latency
stays constant while only the duration increases as we increase the domain. For
all-at-once migration, both latency and duration increase.

We conclude that fluid and batched migration offer a way to bound the
latency impact on a computation during a migration while increasing the
migration duration, whereas all-at-once migration does not.

\subsubsection{Throughput versus processing latency}
\label{sec:eval:tp}

\added{
In this experiment, we evaluate what throughput \pname can sustain for specific latency targets.
As we increase the offered load, we expect the steady-state and migration latency to increase.
For a spcific throughput target, we expect the all-at-once migration strategy to show a higher latency than batched, which itself is expected to be higher than fluid.}

\added{To analyze the latency, we keep the number of keys and bins constant, at \num{16384e6} and 4096, and vary the offered load from \num{250e3} to \num{32e6} in powers of two.
We measure the maximum latency observed during both steady-state and migration for each of the three migration strategies described earlier.
}

\begin{figure}[t]
    \centering
    {\includegraphics[width=.92\linewidth,trim={0 18pt 0 14pt}]{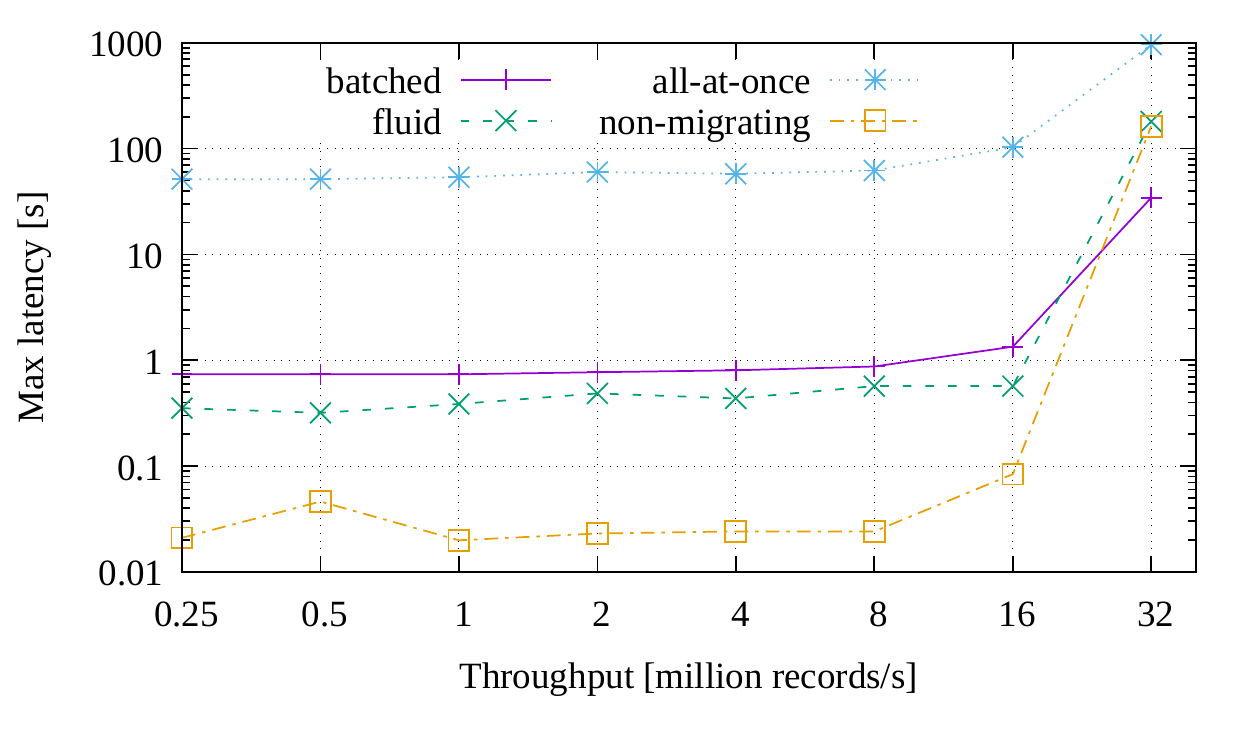}}
    \caption[Latency for different offered rates.]{
\added{
    Offered load versus max latency for different migration strategies for key-count.
    The migration is invariant of the rate up to 16 million records per second.
    }
    }
    \label{fig:keycount-migration-throughput}
\end{figure}

\added{
\Cref{fig:keycount-migration-throughput} shows maximum latency observed when the system is sustaining a certain throughput.
All three migration strategies and non-migrating show a similar pattern: Up to \num{16e6} records per second they do not show a significant increase in latency.
At \num{32e6}, the latency increases significantly, indicating that the system is now overloaded.
}

\added{
We conclude that the system's latency is mostly throughput-in\-vari\-ant until the system saturates and eventually fails to keep up with its input. Both fluid and batched migration sustain a throughput of up to \num{4e6} per second for a latency target of \SI{1}{\second}: \pname's migration strategies can satisfy latency targets 10-100x lower than all-at-once migration with similar throughput.
}

\subsubsection{Memory consumption during migration}\label{sec:eval:mem}

In Section~\ref{sec:eval:migr_const} we analyzed the behavior of different migration strategies when increasing the total amount of state in the system while leaving the state per bin constant.
Our expectation was that the all-at-once migration strategy would always offer the lowest duration when compared to batched and fluid migrations.
Nevertheless, we observe for large amounts of data being migrated the duration for a all-at-once migration is longer than for batched migration.

\begin{figure}[t]
    \centering
    {\includegraphics[width=.92\linewidth,trim={0 14pt 0 8pt}]{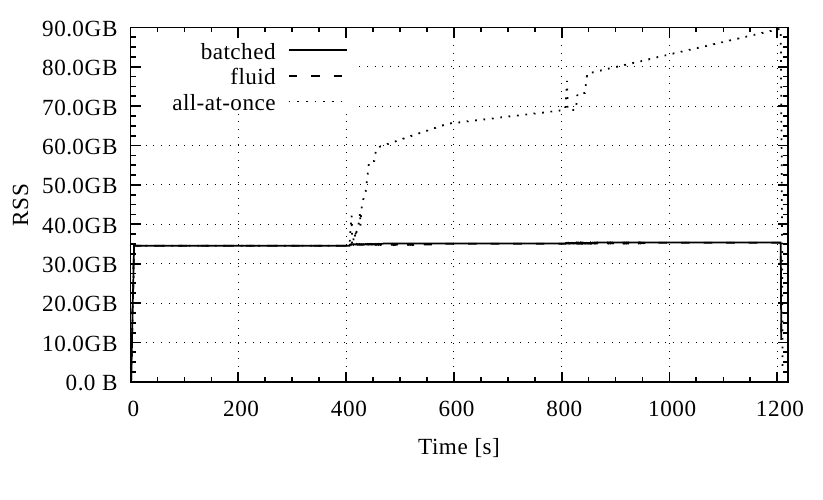}}
    \caption{Memory consumption of key-count per process over time for
    different migration strategies. The fluid and batched strategies require less additional memory in each migration step than the all-at-once migration, which migrates all state at once.}
    \label{fig:wc_migration_mem}
\end{figure}

To analyze the cause for this behavior we compared the memory consumption for
the three migration strategies over time.
We run the key count dataflow with \num{16e9} keys and 4096 bins.
We record the \gls{rss} as reported by Linux over time per process.

Figure~\ref{fig:wc_migration_mem} shows the \gls{rss}
reported by the first timely process for each migration strategy. Batched and fluid
migration show a similar memory consumption of \SI{35}{\gibi\byte} in steady
state and do not expose a large variance during migration at times \SI{400}{\second} and \SI{800}{\second}.
In contrast to that, all-at-once migration shows significant allocations of
approximately additional \SI{30}{\gibi\byte} during the migrations.

The experiment gives us evidence that a all-at-once migration causes significant memory
spikes in addition to latency spikes. The reason for this is that during a
all-at-once migration, each worker extracts and serializes the data to be migrated and enqueues
it for the network threads to send. The network thread's send capacity
is limited by the network throughput, limiting the throughput at which data can
be transferred to the remote host. Batched and fluid migration patterns only
perform another migration once the previous is complete and thus provide a
simple form of flow-control effectively limiting the amount of temporary state.

\balance

\section{Conclusion}

We presented the design and implementation of \pname, which provides efficient, minimally disruptive migration for stream processing systems.
\pname plans fine-grained migrations using the logical timestamps of the stream processor, and interleaves the migrations with regular streaming dataflow processing.
Our evaluation on realistic workloads shows that migration disruption was significantly lower than with prior all-at-once migration strategies.


We implemented \pname in timely dataflow, without any changes to the host dataflow system.
\pname demonstrates that dataflow coordination mechanisms (timestamp frontiers) and dataflow channels themselves are sufficient to implement minimally disruptive migration. \pname's source code is available on \url{https://github.com/strymon-system/megaphone}.


\pagebreak
\bibliographystyle{abbrv}
\bibliography{references}


\end{document}